\documentclass[twocolumn,showpacs,preprintnumbers,footinbib,prl,superscriptaddress,10pt,nofootinbib]{revtex4-1}
\usepackage{placeins}
\usepackage{color}
\usepackage{calc}
\usepackage{amsmath,amssymb,graphicx}
\usepackage{tensor}
\usepackage{bm}
\usepackage{times}
\usepackage[varg]{txfonts}
\usepackage[colorlinks, pdfborder={0 0 0}]{hyperref}
\usepackage{float}
\usepackage{dcolumn}
\usepackage[nolist,nohyperlinks]{acronym}
\usepackage{xspace}
\usepackage[abs]{overpic}
\usepackage{pict2e}
\usepackage{enumitem}
\usepackage[usenames,dvipsnames]{xcolor}
\usepackage[utf8]{inputenc}
\usepackage{acronym}
\usepackage{gensymb}
\usepackage{cleveref}
\usepackage[normalem]{ulem}
\usepackage{longtable}
\usepackage{multirow}
\usepackage{lipsum}
\usepackage{mathtools}
\usepackage{cleveref}

\emergencystretch=\maxdimen
\usepackage{subfigure}

\usepackage{placeins}

\usepackage{booktabs}

\newcommand{\be}{\begin{equation}}
\newcommand{\ee}{\end{equation}}

\newcommand{\msun}{{\rm M}_{\odot}}

\begin{document}
 
 \title{Detectable environmental effects in GW190521-like black-hole binaries with LISA}

\author{Alexandre Toubiana}
\affiliation{APC, AstroParticule et Cosmologie, 
Université de Paris, CNRS,  F-75013 Paris, France}
\affiliation{Institut d'Astrophysique de Paris, CNRS \& Sorbonne
 Universit\'es, UMR 7095, 98 bis bd Arago, 75014 Paris, France} 

\author{Laura Sberna}
\affiliation{Perimeter Institute, 31 Caroline St N, Ontario, Canada} 

\author{Andrea Caputo}
\affiliation{Instituto de Fisica Corpuscular, Universidad de Valencia and CSIC, Edificio Institutos Investigacion, Catedratico Jose Beltran 2, Paterna, 46980 Spain}
\affiliation{School of Physics and Astronomy, Tel-Aviv University, Tel-Aviv 69978, Israel}\affiliation{ Department of Particle Physics and Astrophysics,
Weizmann Institute of Science, Rehovot 7610001,Israel}

\author{Giulia Cusin}
\affiliation{Universit\'e de Gen\'eve, D\'epartement de Physique Th\'eorique and Centre for Astroparticle Physics, 24 quai Ernest-Ansermet, CH-1211 Gen\'eve 4, Switzerland}

\author{Sylvain Marsat}
\affiliation{APC, AstroParticule et Cosmologie, 
Université de Paris, CNRS,  F-75013 Paris, France}

 \author{Karan Jani}
 \affiliation{Department Physics and Astronomy, Vanderbilt University, 2301 Vanderbilt Place, Nashville, TN, 37235, USA }

\author{Stanislav Babak}
\affiliation{APC, AstroParticule et Cosmologie, 
Université de Paris, CNRS, F-75013 Paris, France}
\affiliation{Moscow Institute of Physics and Technology, Dolgoprudny, Moscow region, Russia}
 
\author{Enrico Barausse}
\affiliation{SISSA, Via Bonomea 265, 34136 Trieste, Italy \& INFN,
Sezione di Trieste}
\affiliation{IFPU - Institute for Fundamental Physics of the Universe,
Via Beirut 2, 34014 Trieste, Italy}

\author{Chiara Caprini}
\affiliation{APC, AstroParticule et Cosmologie, 
Université de Paris, CNRS, F-75013 Paris, France}

\author{Paolo Pani}
\affiliation{Dipartimento di Fisica, ``Sapienza'' Universit\`a di Roma \& Sezione INFN Roma1, Piazzale Aldo Moro 5, 00185, Roma, Italy}

\author{Alberto Sesana}
\affiliation{Department of Physics G. Occhialini, University of Milano - Bicocca, Piazza della Scienza 3, 20126 Milano, Italy}
\affiliation{National Institute of Nuclear Physics INFN, Milano - Bicocca, Piazza della Scienza 3, 20126 Milano, Italy}

\author{Nicola Tamanini}
\affiliation{Max-Planck-Institut für Gravitationsphysik, Albert-Einstein-Institut, Am Mühlenberg 1,14476 Potsdam-Golm, Germany.}

 \begin{abstract}
  GW190521 is the compact binary with the largest  masses observed to date, with at least one black hole in the pair-instability gap. This event has also been claimed to be  associated with an optical flare observed by the Zwicky Transient Facility in an Active Galactic Nucleus (AGN), possibly due to the post-merger motion of the merger remnant in the  AGN gaseous disk.  The Laser Interferometer Space Antenna (LISA) may detect up to ten of such gas-rich black hole binaries months to years before their detection by LIGO/Virgo-like interferometers, localizing them in the sky within $\approx1$ deg$^2$. LISA will also measure directly deviations from purely vacuum and stationary waveforms, arising from gas accretion, dynamical friction, and orbital motion 
  around the AGN's  massive black hole
  (acceleration,  strong lensing, and Doppler modulation).
LISA will therefore be crucial to alert and point electromagnetic telescopes ahead of time on this novel class of gas-rich sources, to gain direct insight on their physics, and to disentangle environmental effects from corrections to General Relativity that may also appear in the waveforms at low frequencies.
 \end{abstract}
    
 \maketitle

\noindent 
  GW190521 is the most massive compact binary merger observed by the LIGO/Virgo Collaboration (LVC), with progenitor black-hole (BH) masses of
$85^{+21}_{-14}\,M_\odot$ and $66^{+17}_{-18}\, M_\odot$~\cite{Abbott:2020tfl, Abbott:2020mjq}. The larger  BH lies in the pair-instability gap $\sim[50,130] M_\odot$ \cite{2002RvMP...74.1015W,2003ApJ...591..288H,2019ApJ...887...53F}, calling for interpretations beyond standard stellar-evolution models. 
A viable channel to produce such massive BHs
is via repeated mergers
(which would also explain the large misaligned spins of 
GW190521~\cite{Abbott:2020tfl, Abbott:2020mjq}), e.g. in stellar clusters \cite{2017PhRvD..95l4046G,2019PhRvD.100d3027R,2020arXiv200905065F}
or  active galactic nuclei~(AGNs)~\cite{2007MNRAS.374..515L,2020MNRAS.494.1203M,2020ApJ...898...25T}. While repeated mergers might be rare in globular clusters, due to BH ejection by gravitational recoil~\cite{Abbott:2020mjq}, nuclear star clusters  have higher escape velocities and  more efficiently retain merger remnants. Alternatively,  
GW190521 may have formed in an AGN disk,
where mass segregation and dynamical friction~(DF) favor BH accumulation near the center (enhancing merger rates)
and their growth by mergers and accretion~\cite{2007MNRAS.374..515L,2020MNRAS.494.1203M,2020ApJ...898...25T,Bartos:2016dgn,Yang:2019okq}. The large GW190521 masses  may also be consistent  with metal-free, population-III star progenitors~\cite{2020arXiv200906922K} (see also~\cite{Kinugawa:2020ego,Safarzadeh:2020vbv,Farrell:2020zju}).
Other less standard scenarios include beyond-Standard-Model physics~\cite{Sakstein:2020axg}, primordial BHs~\cite{DeLuca:2020sae}, boson stars \cite{2020arXiv200905376C}, and extensions of General Relativity~(GR)~\cite{2020arXiv200904360M}. Finally, \cite{2020arXiv200905472F,Nitz:2020mga} note that there is a non-negligible probability of GW190521 being a ``straddling'' binary,
with  components below and above the pair-instability gap. Although some analyses~\cite{2020arXiv200905461G,2020arXiv200904771R,CalderonBustillo:2020odh} suggest that  eccentric waveforms might fit the data better, supporting a dynamical origin in dense environments, the formation of GW190521 remains mysterious.

Remarkably, the Zwicky Transient Facility  observed an optical flare (ZTF19abanrhr), interpreted as coming from the kicked GW190521 BH merger remnant moving in an AGN disk~\cite{PhysRevLett.124.251102}. If confirmed, this would be the first  electromagnetic counterpart to a BH coalescence (see, however, \cite{Connaughton:2016umz,Savchenko:2016kiv}).
The flare  occurred $\sim34$ days after GW190521
(the delay being ascribed to the remnant's recoil)
in AGN J124942.3+344929 at redshift $z = 0.438$. If the flare is indeed associated with GW190521
and due to the remnant's recoil in the AGN disk,
\cite{PhysRevLett.124.251102}  finds a total binary mass $\sim 150 M_\odot$, kick velocity $\sim 200$ km/s at $\sim 60$ deg from the disk's midplane, a disk aspect ratio (height to galactocentric radius) of $H/a\sim 0.01$, and a gas density of $\rho \sim 10^{-10}$ g/cm$^3$. The~\cite{PhysRevLett.124.251102} authors also argue that the binary is most likely located in a disk migration trap (galactocentric distance $a\sim 700 GM/c^2$, with $M\sim10^8 -10^9 \ M_{\odot}$  the mass of the AGN's  BH), where gas torques
vanish and binaries  accumulate as they migrate inward \cite{2016ApJ...819L..17B}.
While
the GW190521--ZTF19abanrhr association is debated~\cite{Ashton:2020kyr},
 our results   will not  rely on it, but only assume that GW190521-like systems reside in gas-rich environments (e.g. AGNs~\cite{2007MNRAS.374..515L,2020MNRAS.494.1203M,2020ApJ...898...25T,Bartos:2016dgn,Yang:2019okq}).

Months or years before merging in the 
LIGO/Virgo  band, BH binaries with masses above a few tens $M_\odot$ spiral in the mHz band of the Laser Interferometer Space Antenna (LISA)~\cite{Audley:2017drz}, a  gravitational-wave~(GW) space-borne
experiment scheduled for 2034.
 Observing the inspiral with LISA  would permit estimating the source parameters very precisely~\cite{Sesana:2016ljz, 2020NatAs...4..260J, Toubiana:2020cqv}, e.g., the chirp mass and  distance to fractional errors $\sim 10^{-4}$ and $\sim 0.4$, respectively, the sky position below $\sim 1$ deg$^2$~-- and it would allow for predicting the coalescence time within a minute, weeks before the signal is detected from Earth.
 This would permit alerting electromagnetic telescopes in advance, pointing them at smaller sky regions, and looking for  electromagnetic counterparts~\cite{Sesana:2016ljz} {\it coincident} with the coalescence. Moreover, GR extensions typically predict low-frequency corrections to the GW phase, e.g. vacuum dipole emission at $-1$ post-Newtonian~(PN) order, which will be tested to exquisite precision by LISA inspiral observations~\cite{Barausse:2016eii,Vitale_2016, Toubiana:2020vtf}.

We argue below  that LISA might detect several gas-rich, high-mass BH binaries. Besides observing these sources  beforehand and localizing them accurately \cite{PRD}, LISA will also detect the environmental (i.e. non-vacuum~\cite{Barausse:2014tra}) effects  in the GW signal, namely gas accretion and DF on the component BHs, the binary's  acceleration around the AGN's BH, and possibly the Doppler modulation
and the lensing/Shapiro time delay  from the central BH. These effects may
be degenerate with low-frequency tests of GR, but may help localize the source by correlating with AGN catalogs.
 Henceforth, we use units where $G=c=1$.
 
\noindent 
 \textit{Event rates.} Assuming  binaries  with parameters drawn from the LVC posteriors, Ref.~\cite{Abbott:2020mjq}  estimates the comoving merger rate of GW190521-like systems as $0.13^{+0.30}_{-0.11}$ yr$^{-1}$ Gpc$^{-3}$. With the same hypotheses, LISA will observe  1--10 such systems, depending on  high-frequency laser noise,  mission lifetime, and  operation duty cycle, for $z\lesssim 0.5$ (for more details see~\cite{PRD}).
 However, if GW190521 lies at the low-mass end of a heavy-BH population extending beyond $100\msun$, 
  LISA rates
 would be significantly higher (because in the $(10-10^3)\msun$ range  the LISA horizon distance  is   $\propto{\cal M}^{5/3}$, with ${\cal M}$ the source-frame chirp mass~\cite{2020NatAs...4..260J}). Indeed,  conservatively assuming a Salpeter mass function extending to $200\msun$, the inferred LVC rate would boost LISA detections by about an order of magnitude (see also \cite{Ezquiaga:2020tns}).

\noindent 
\textit{Detectability of environmental effects.} We model accretion from the AGN  disk  by the Eddington ratio $f_{\rm Edd}\equiv \dot{m}/\dot{m}_{\rm Edd}$ between the
mass accretion rate $\dot{m}$ of either BH and
$\dot{m}_{\rm Edd}\equiv L_{\rm Edd}/\eta$, with $ L_{\rm Edd}$ the Eddington luminosity and $\eta\approx 0.1$ the radiative efficiency. The mass growth of the BHs, ($i=1,2$), is 
$m_i(t) = m_{i}(0) \, \exp(f_{\rm Edd} \,  t  /\tau_S)$,
with $ \tau_S= 4.5\times 10^{7}\, {\rm yr}$ the Salpeter time.
The phase term in the Fourier-domain GW signal $\tilde{h} \sim |\tilde{h}| \, e^{i \tilde\phi} $ induced by the mass growth can  be evaluated in the stationary-phase approximation at leading PN order~\cite{Caputo:2020irr}:
\begin{equation}
    \label{phaseA}
\tilde\phi_{\rm accretion} \approx  - f_{{\rm Edd}} \,  (8 \, \xi +15) \, \frac{ 75  \, \mathcal{M}  }{851 \ 968 \,  \tau_S } \, \left[\pi  f 
\mathcal{M} (1+z)\right]{}^{-13/3} \, 
,
\end{equation}
with $f$ the observed GW frequency, $z$ the redshift at coalescence, and $\xi\sim{\cal O}(1)$ a factor parameterizing the drag  produced by momentum transfer from the accreted gas~\cite{Caputo:2020irr}, which we conservatively set to zero. 

The binary's center of mass (CoM)  acceleration around the central massive BH also
modifies the  waveform (c.f.~\cite{Bonvin:2016qxr,Inayoshi:2017hgw,Tamanini:2019usx} for CoM~accelerations almost constant during the observation period).
The phase correction reads~\cite{Bonvin:2016qxr,Inayoshi:2017hgw,Tamanini:2019usx} %
\begin{equation}
	\tilde\phi_{\rm acceleration} \approx
	 \frac{25\, \mathcal{M}}{65 \ 536}  \dot{v}^\shortparallel(t_{c})\, \left[\pi f
	 \mathcal{M}(1+z)\right]^{-13/3} \,,
	 \label{phaseACCEL}
\end{equation}
where $\dot{v}^\shortparallel$ is the acceleration  along the line of sight,
computed at coalescence,
dominating over the cosmological acceleration (which we neglect). For quasicircular galactocentric orbits, 
$\dot{v}^\shortparallel
\approx (3.2\times 10^{-11}{\rm m/s}^2)\,\epsilon$, where \cite{Bonvin:2016qxr}
\begin{equation}\label{eq:epsilon}
\epsilon=\left(\frac{v_{\rm orb}}{100\, {\rm km}/{\rm s}}\right)^2\frac{10\,{\rm kpc}}{a}\cos\psi\,,
\end{equation}
with $v_{\rm orb}$ the galactocentric orbital velocity, and
$\psi$ the angle between  line of sight and  acceleration.
Since $\cos\psi=\cos{\iota}\sin(\Omega t+\phi_0)$ (with $\iota$ the inclination angle of the line of sight relative to the AGN disk, $\Omega=\sqrt{M}/a^{3/2}$,  and $\phi_0$ the initial phase),
the assumption of constant acceleration only holds 
at sufficiently large galactocentric distances $a$  (i.e. low $\Omega$). We will verify this assumption {\it a posteriori} (and relax it) later.

Eqs.~\eqref{phaseA}--\eqref{phaseACCEL} show that  accretion and (constant) acceleration are degenerate, since they both appear at  $-4$PN order~\cite{Barausse:2014tra,Cardoso:2019rou,Bonvin:2016qxr,Caputo:2020irr} (accretion always yields a negative phase contribution, while acceleration can give contributions of either sign). Both effects can be included in the waveform via a phenomenological PN term $\tilde\phi_{\rm -4PN}=\varphi_{-4} \left[\pi  f \mathcal{M}(1+z)\right]{}^{-13/3}$~\cite{TheLIGOScientific:2016src}, with $\varphi_{-4}$ related to $f_{\rm Edd}$ and $\epsilon$ for accretion and acceleration, respectively.


We also consider the DF from  gas with density $\rho$ surrounding the binary.
Assuming a binary's CoM approximately comoving with the gas, the DF exerts a drag force on each BH (opposite to the BH's velocity $\vec{v}_i$ in the CoM frame), $F_{{\rm DF},i} = {4\pi \rho (G m_i)^2} I(r_i,v_i)/{v_i^2}$, where $r_i$ is the distance of the BH from the CoM, and we  assume $v_i\gg c_s$ (with $c_s$ the sound speed; note that $v_i$ is relativistic 
for binaries in the LISA band). We use the analytic  ``Coulomb logarithm'' $I(r,v)$ of~\cite{Kim:2007zb}, which  was validated against simulations for $v_i/c_s\lesssim8$, but  which we extrapolate further (c.f. \cite{Kocsis:2011dr}).

We assume  $c_s\approx v_{\rm orb} (H/a)$~\cite{2002apa..book.....F}, with $H/a\sim 0.01$~\cite{PhysRevLett.124.251102}.
Following~\cite{Kim:2007zb} (see also \cite{Ostriker:1998fa,Barausse:2007ph,Macedo:2013qea}), we only include
 the effect of the wake created by each BH on itself, and neglect the companion's~\cite{Kim:2008ab}. For $f \lessapprox 0.3\,{\rm Hz}$ this is a good approximation, since the orbital separation of GW190521 in the LISA band is larger than the wake's size. 
In the adiabatic approximation,
the DF-induced phase correction  first enters 
 at $-5.5$PN order:    
\begin{equation}
    \tilde\phi_{\rm DF} \simeq   - \rho \frac{ 25 \pi  (3 \nu -1) \mathcal{M}^2}{739 \ 328 \, \nu ^2}  \, \gamma_{\rm DF}
    \left[\pi f \mathcal{M}(1+z)\right]^{-16/3}\,, \label{phaseDF}
\end{equation}
with $\gamma_{\rm DF} = -247 \log \left( f /f_{\rm DF}\right)-39+304 \log (20)+38 \log \left({3125}/{8}\right)$ and $f_{\rm DF}=  {c_s}/{[22 \pi (m_1+m_2) ]} $, being $\nu = m_1 m_2/(m_1+m_2)^2$ the symmetric mass ratio.

\begin{figure}[!ht]
\includegraphics[width=1\columnwidth]{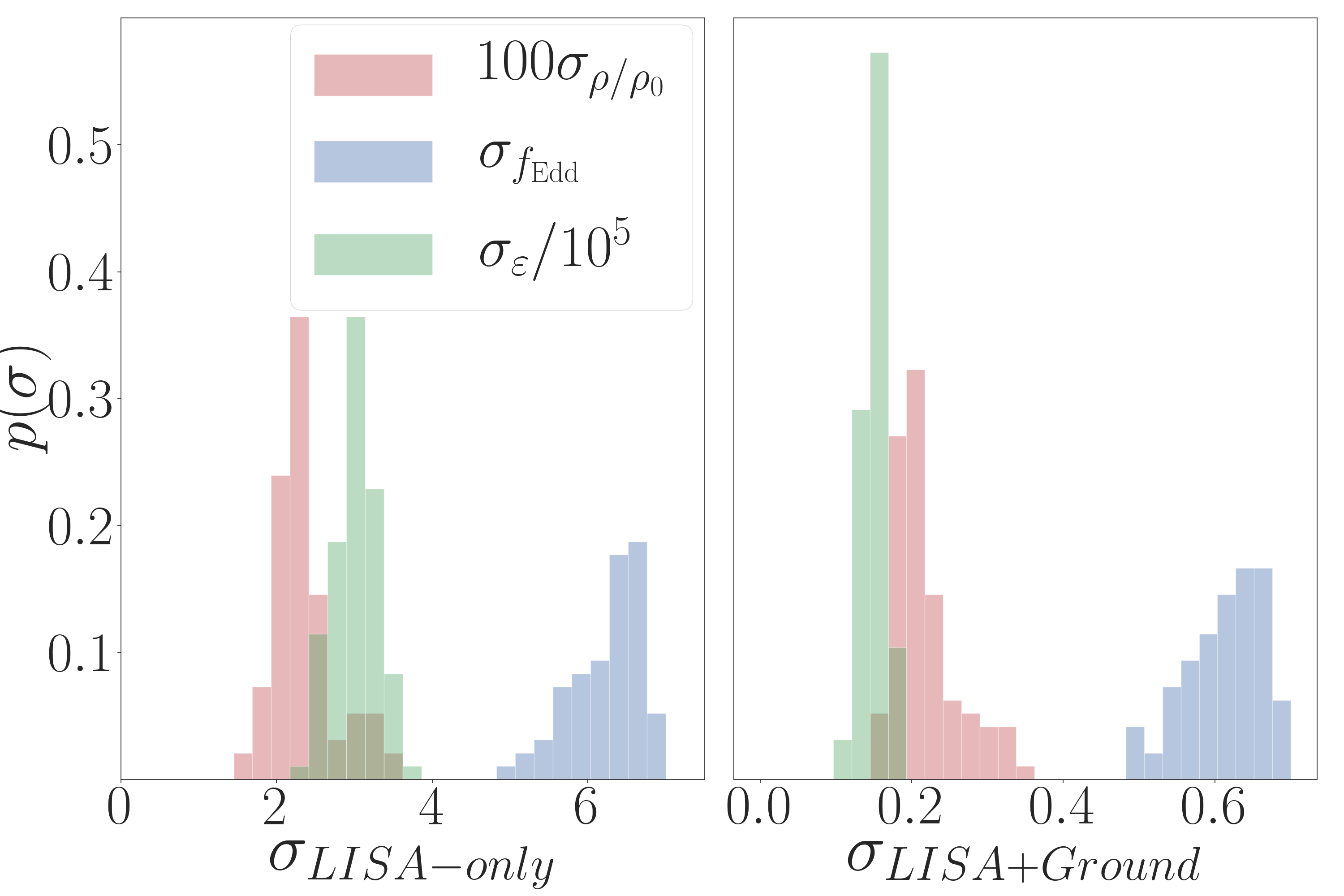}
\caption{Distribution of Fisher-matrix errors on environmental effects for the LVC samples with LISA alone or jointly with ground detectors.} \label{errors_samples}
\end{figure}

In Fig.~\ref{errors_samples}, we show the distribution of errors 
-- produced with the augmented Fisher formalism of \cite{Toubiana:2020cqv} -- for the Eddington rate $f_{\rm Edd}$, the acceleration parameter $\epsilon$, and the gas density $\rho$,
(normalized to $\rho_0=10^{-10}\,{\rm g\, cm}^{-3}$~\cite{PhysRevLett.124.251102}), considering  modifications one by one. We use
$f_{\rm Edd}=\rho=\epsilon=0$ as injections, i.e.~the distributions represent  optimistic upper bounds  on the parameters. We use the LVC masses, spins, distance and inclination samples for the NRSur7dq4 model~\cite{Varma:2019csw} and, for each sample, we set the time to coalescence  to the mission's duration (6 years) and draw sky location and polarization randomly. The extra terms of Eqs.~\eqref{phaseA}, \eqref{phaseACCEL}, and~\eqref{phaseDF} were added to PhenomD waveforms~\citep{PhysRevD.93.044006,KHH16} one by one, and we
accounted for the antenna motion  during  observations following~\cite{Marsat:2018oam,Marsat:2020rtl,Toubiana:2020cqv}. 
We consider detections by LISA alone, and
jointly with ground interferometers. In the \emph{LISA+Ground} case, we mimic a multiband detection by assuming that masses, spins, and merger time are measured by ground detectors, thus removing them from the analysis.

We find that LISA alone
can detect super-Eddington accretion rates ($f_{\rm Edd}\gtrsim6$), which may be typical in dense environments~\cite{Li:2019hfq}, and  acceleration parameters $\epsilon\gtrsim 3\times 10^5$,  corresponding to 
$a\approx$ 1 pc for $M= 10^8 M_\odot$.
The DF effect is even stronger, with $\rho/\rho_0$ constrained at  percent level. All errors improve by about an order of magnitude in the \emph{LISA+Ground} scenario (e.g., sub-Eddington accretion rates
become measurable).

Next, we
focus on one system compatible with the LVC posteriors and perform a Markov-Chain-Monte-Carlo analysis~\cite{Karandikar2006} like in~\cite{Toubiana:2020cqv}, injecting nonzero values (plausible for sources in AGNs) for all  environmental effects (considered simultaneously): $f_{\rm Edd}=5$, $\epsilon=3.2\times 10^6$ (corresponding to $a\approx  0.4 \ {\rm pc}$ for $M=10^8 \ M_{\odot}$) 
and $\rho=\rho_0$.  
 \begin{figure}[!t]
\includegraphics[width=0.9\columnwidth]{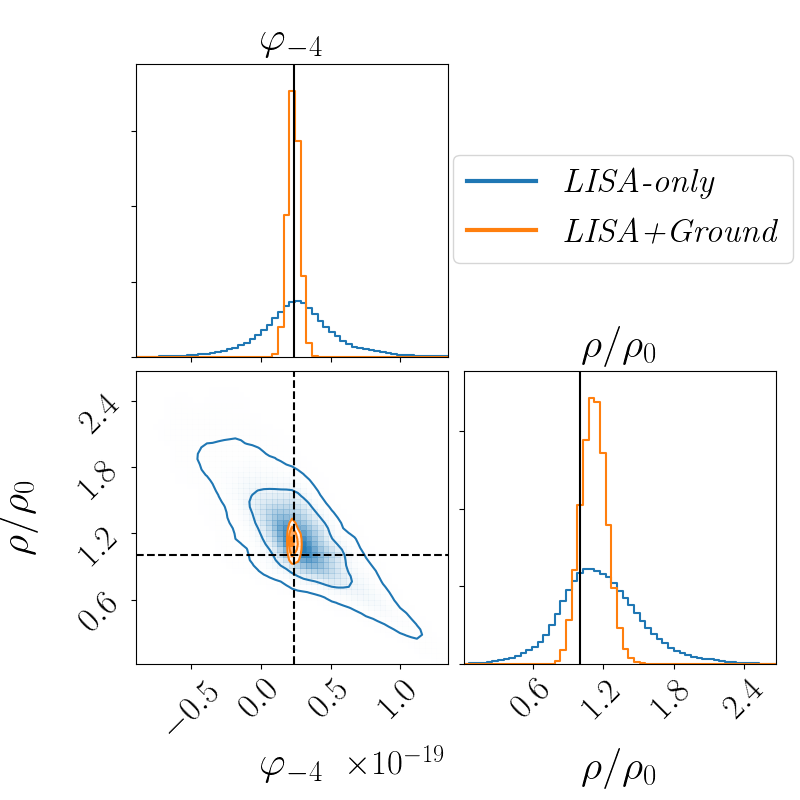}
\caption{Posterior distribution of gas density and -4PN phase term (i.e. constant acceleration and/or accretion), with  $68 \%$, $95\%$ and $99\%$ confidence contours for a best-case event consistent with GW190521. Black lines indicate the injected values.}\label{MoneyPlot2}
\end{figure} 
Fig.~\ref{MoneyPlot2}  shows the  posterior distributions for the density ($\rho/\rho_0$) and the parameter $\varphi_{-4}$ accounting for acceleration and accretion.
Both parameters can be measured well, since they appear at different (negative) PN orders. Note that the sign of $\varphi_{-4}$ can help distinguish 
accretion ($\varphi_{-4}<0$) from  acceleration ($\varphi_{-4}$ of either sign). We focus on a system with a (LISA-only) signal-to-noise ratio (SNR) of 9.6, the highest among the LVC events, at a distance $\sim 1.4 \ {\rm Gpc}$. Since this system is close to the LISA detection threshold  (${\rm SNR}\sim8$), we conclude that 
even for near-threshold events environmental effects are measurable, and their 
observation will only be limited by the event  detectability.

\begin{figure*}[!ht]
\includegraphics[width=2.\columnwidth]{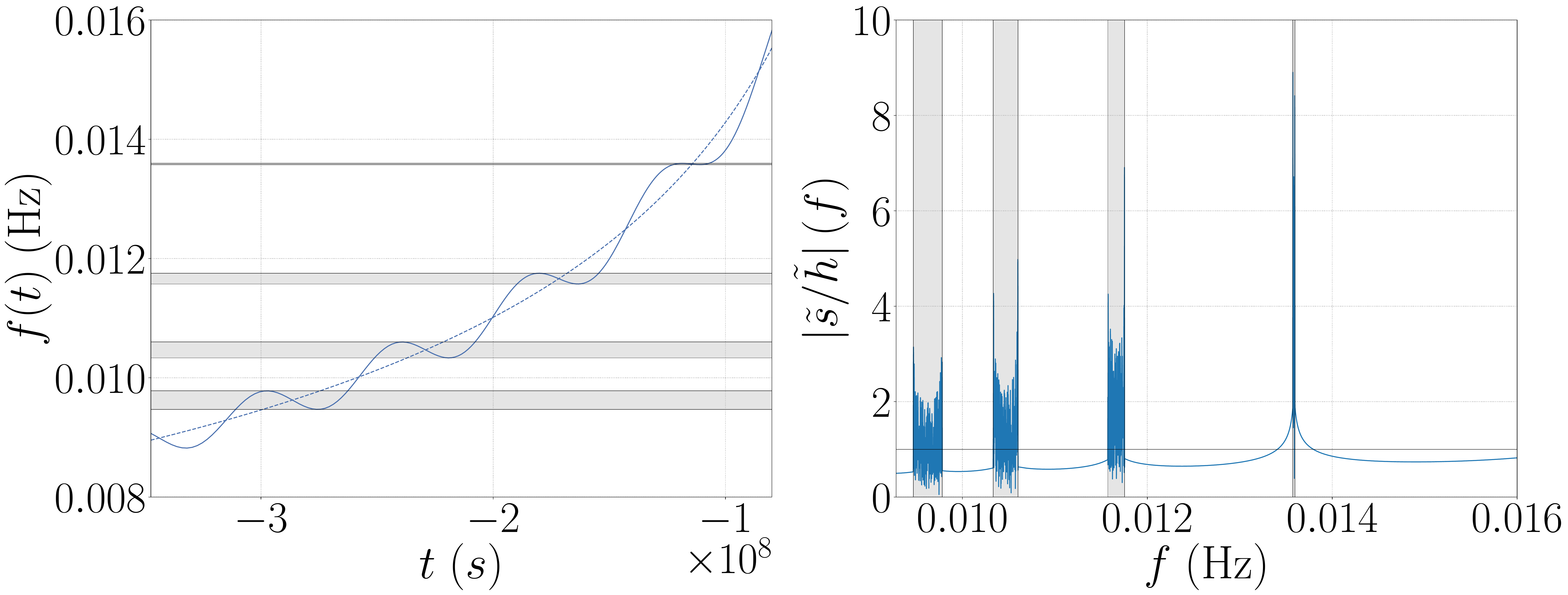}
\caption{Doppler modulations of the GW signal due to the motion around the central BH. The left panel shows the time-frequency track of the modulated signal (solid line) compared to the nonmodulated one (dashed) (coalescence is at $t=0$). The right panel shows the amplitude of the Fourier-domain transfer function $(\tilde{s}/\tilde{h}) (f)$, with a horizontal line at 1. In both panels, the shaded bands show the frequency bands where the time-to-frequency map becomes multivalued.}\label{TF_GW}
\end{figure*} 

We  have checked that the assumption of constant acceleration holds, i.e.~verified that the systematic error produced by the variation of $\epsilon$ over the observation time $T_{\rm obs}$ is negligible relative to the statistical error. 
For $a\lesssim 0.25\, {\rm pc}\, [M/(10^8 M_\odot)]^{3/7}[T_{\rm obs}/(6 {\rm yr})]^{2/7}$ (i.e.~orbital periods $T\lesssim 1200\,{\rm yr}\, [M/(10^8 M_\odot)]^{1/7}[T_{\rm obs}/(6 {\rm yr})]^{3/7}$), however, this may no longer be true. This is the case, e.g.,~if GW190521 lies in a disk migration trap. Ref.~\cite{PhysRevLett.124.251102} estimates the trap's distance from the central BH as $a\sim 700 M$,
corresponding to $T\sim 1.8\,{\rm yr}$, i.e. the acceleration cannot be assumed  constant over the observation time. In this situation, the Taylor expansion leading to Eq.~\eqref{phaseACCEL} breaks down, and the GW signal can be estimated as $s(t) = h(t+d^\shortparallel(t))$,
with $h(t)$ the source-frame strain. The delay 
$d^\shortparallel(t)$ arises from the change in the source distance 
due to the orbital motion, and is given by the orbit's projection on the line of sight: $ d^\shortparallel(t) = a \cos \iota \sin(\Omega t + \phi_0)$.
This time-varying delay produces an oscillating Doppler modulation $\phi_{\rm Doppler} \sim 2\pi f d^\shortparallel$ of the observed  signal. The magnitude of this phase is approximately
\begin{equation}
    \label{eq:OscDoppler}
    2\pi f a \sim  2\times 10^4\, {\rm rad} \left(\frac{M}{10^8 M_\odot}\right) \left(\frac{f}{10\, {\rm mHz}}\right) 
    \left( \frac{a}{700 M} \right)\,.
\end{equation} 
This effect  strongly impacts the signal, dominating the Doppler modulation produced by the LISA motion ($\approx 30\, {\rm rad}$), and happening on comparable  timescales.
The GW frequency suffers redshifts or blueshifts as the binary's CoM moves away from or toward LISA; see Fig.~\ref{TF_GW}. These modulations  dominate the GW-driven chirp rate, leading to a
multivalued time-to-frequency map in the shaded bands of Fig.~\ref{TF_GW}, where the chirping and anti-chirping parts of the signal overlap. This strongly affects the Fourier-domain observed signal $\tilde{s}(f)$, with the transfer function amplitude $|\mathcal{T}(f)| = |\tilde{s}(f)/\tilde{h}(f)|$ showing interference patterns in the shaded  bands. The impact on 
 detection and parameter estimation is under study~\cite{PRD}.

\begin{figure}[!ht]
\includegraphics[width=0.9\columnwidth]{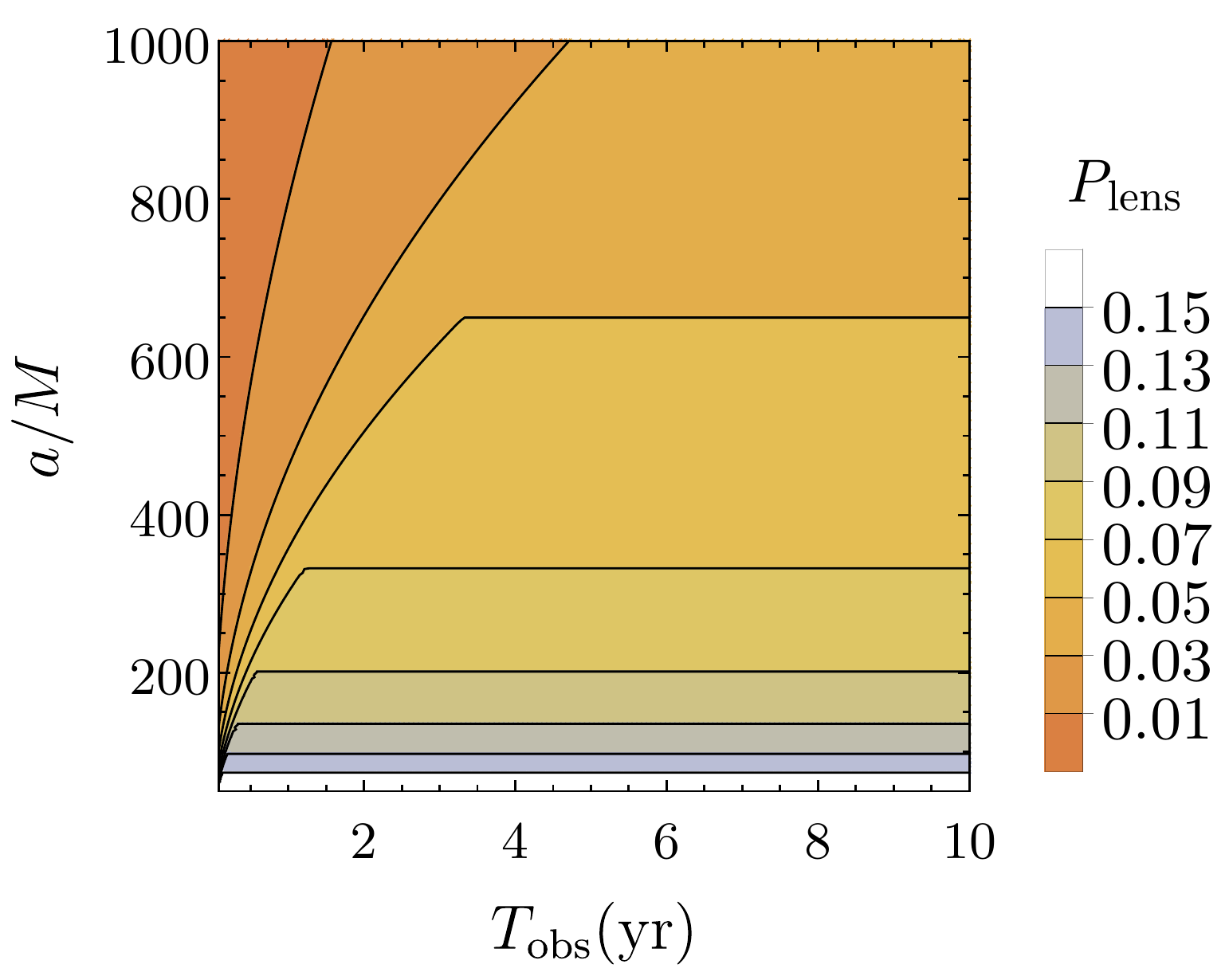}\quad
\caption{Lensing probability  as  function of observation time, $T_{\rm obs}$, and galactocentric distance, $a/M$, for $M=10^8 M_{\odot}$.\label{Prob3}}
\end{figure}

Another potentially detectable effect, particularly for edge-on AGN disks, is the strong lensing of the GWs by the central BH, which occurs at scales given by the Einstein radius
\be\label{rE}
r_E\simeq (4 M D_{A})^{1/2}=[4 M a \cos{\iota} \sin(\Omega t+\phi_0)]^{1/2}\,,
\ee
where we assume $a \ll D_A$, with $D_A$ the lens angular diameter distance. 
Significant lensing occurs when the source passes within $\sim r_E$ of the lens, and the lensing probability 
is thus the fraction of time (during a full orbit around the central BH) for which this happens~\cite{DOrazio:2019fbq,2018MNRAS.474.2975D}. A GW190521-like event in an AGN disk's migration trap falls  either in the repeating-lens regime or in  the slowly moving lens regime defined e.g. in~\cite{2018MNRAS.474.2975D}, depending on the observation time $T_{\text{obs}}$ and $M$. The probability of strong lensing is (see Fig.~\ref{Prob3}) 
\begin{align}
P_{\text{lens}}
&=\text{Min}\left[ \frac{10^{12}}{2} \frac{T_{\rm obs}}{\text{yr}}\left(\frac{a}{M}\right)^{-3/2}\frac{M_{\odot}}{M},1\right]\frac{2}{\pi}\arcsin\left[2 \sqrt{\frac{M}{a}} \right]\,.
\end{align}
. 

Strong lensing also affects the observed waveform directly. For a  plane wave, the lensed signal (in real space) is given by
\be\label{hL}
h^L(t)=F(f, t)\, h(t)\,,
\ee
in terms of the amplification factor $F(f,t)$. For a pointlike lens in the geometric-optics approximation~\cite{1992grle.book, Takahashi:2003ix},
\be\label{amplification}
F(f, t)=|\mu_{+}|^{1/2}-i|\mu_{-}|^{1/2}e^{2\pi i f \Delta t}\,,
\ee
where the magnification of each image,  $\mu_{\pm}=1/2 \pm  (y^2+2)/(2y\sqrt{y^2+4})$, depends on time  through the lensing parameter $y\equiv b/r_E$, with $b$  the impact parameter. 
The time delay between  two images is
$\Delta t=\Delta t_{\text{fid}}\left[y/2\sqrt{y^2+4}+\ln\left((\sqrt{y^2+4}+y)/(\sqrt{y^2+4}-y)\right)\right]$ 
where $\Delta t_{\text{fid}}\simeq 2\times 10^{-5} (1+z)M/M_{\odot} \text{sec}$. 
Periodic passages of the orbit behind the central massive BH will produce repeated interference patterns on the observed waveform. From Eq.~\eqref{hL}, one  sees that besides rescaling the waveform amplitude, strong lensing also yields an additive correction to the phase. For a plane wave, the latter is simply $\phi_{\rm SL}=\text{Arg}[F(f, t)]$.
We have checked, however, that this dephasing is typically smaller than the Doppler modulation described above (c.f. \cite{PRD} for details).

\textit{Discussion.} 
If GW190521-like events occur in dense gaseous environments, these GW sources
may provide a unique probe of AGN properties. By detecting the inspiral of these binaries months or years before their coalescence in the band of ground-based detectors~\cite{PRD}, LISA will measure the binary parameters with exquisite precision
and help localize potential electromagnetic counterparts (which would in turn help identify the host galaxy among the thousands -- of which just a  few AGNs -- present within the $1 \ {\rm deg}^2$ LISA error-box). Moreover, LISA
may uncover nonvacuum effects carrying information on the astrophysical source environment, further confirming the identification with an electromagnetic counterpart (if any).

Detecting environmental effects is also crucial to test GR~\cite{Barausse:2014tra}. Effects like extra dimensions or a time-varying Newton constant enter the inspiral waveform at $-4$PN order~\cite{Yunes:2016jcc}, being thus degenerate with accretion or acceleration. Other effects, like vacuum dipole emission,  also appear at negative PN orders~\cite{Barausse:2016eii}.
If negative PN corrections are measured by standard parameterized tests~\cite{TheLIGOScientific:2016src},  an electromagnetic counterpart in the LISA sky-position error box would
favor an environmental origin over 
 a beyond-GR one.
 
Given these tantalizing prospects, dedicated simulations would be needed to carefully describe environmental effects in binaries, 
 accounting e.g. for radiative transfer and outflows that may affect  DF and accretion~\cite{Li:2019hfq, Gruzinov:2019gpd}, and modeling strong accelerations in relativistic binaries accurately~\cite{PRD}. 
Relativistic corrections might also impact the signal, if the 
binary is close to the central BH (note that $a\approx 700 M$ corresponds to $v_{\rm orb}\approx 0.04$). These include orbital periapse precession, spin-orbit~\cite{Yu:2020dlm} and spin-spin~\cite{spin-spin-coupling} precession, and gravitational redshift.
The Shapiro time delay can also be significant, i.e. $\delta t \sim 10^3\, {\rm s}\left(\frac{M}{10^8 M_{\odot}}\right)$ as we will discuss in~\cite{PRD}.
The central BH may also produce Lidov-Kozai oscillations, increasing the binary's eccentricity~\cite{1962P&SS....9..719L,1962AJ.....67..591K}.

Besides the merger remnant's emission in optical, X-rays from accretion onto the binary components and radio flares from jets are expected. Following~\cite{Caputo:2020irr}, we find both effects difficult to observe even with future telescopes (e.g. Athena+, Square Kilometer Array)~\cite{PRD}.

\begin{acknowledgments}
\noindent{{{\em Acknowledgments.}}}
We acknowledge financial support from the European Union's H2020 ERC Consolidator Grants
``GRavity from Astrophysical to Microscopic Scales'', grant agreement no. GRAMS-815673
(to E.B.), and ``Binary Massive Black Hole Astrophysics'', grant agreement no. BMassive-818691 (to A.S.); from the Swiss National Science Foundation; from the French space agency CNES in the framework of the LISA mission;  from the European Union's H2020 ERC, Starting Grant agreement no.~DarkGRA--757480 (to P.P.); 
from 
the MIUR PRIN and FARE programmes (GW-NEXT, CUP:~B84I20000100001, to P.P.).
We thank I. Bartos, J.-M. Ezquiaga, J. Calderon Bustillo, and especially T. Dal Canton for insightful comments.
 \end{acknowledgments}

\bibliography{Ref}
\end{document}